**Frequency-Division Multiplexing in Magnonic Logic Networks Based on Caustic-Like Spin-Wave Beams**

*Frank Heussner[1,*], Matthias Nabinger[1], Tobias Fischer[1,2], Thomas Brächer[1], Alexander A. Serga[1], Burkard Hillebrands[1], and Philipp Pirro[1]*

[1] Fachbereich Physik and Landesforschungszentrum OPTIMAS, Technische Universität Kaiserslautern, D-67663 Kaiserslautern, Germany
[2] Graduate School Materials Science in Mainz, 55128 Mainz, Germany
* E-mail: heussner@rhrk.uni-kl.de

Abstract:

Wave-based data processing by spin waves and their quanta, magnons, is a promising technique to overcome the challenges which CMOS-based logic networks are facing nowadays. The advantage of these quasi-particles lies in their potential for the realization of energy efficient devices on the micro- to nanometer scale due to their charge-less propagation in magnetic materials. In this paper, the frequency dependence of the propagation direction of caustic-like spin-wave beams in microstructured ferromagnets is studied by micromagnetic simulations. Based on the observed alteration of the propagation angle, an approach to spatially combine and separate spin-wave signals of different frequencies is demonstrated. The presented magnetic structure constitutes a prototype design of a passive circuit enabling frequency-division multiplexing in magnonic logic networks. It is verified that spin-wave signals of different frequencies can be transmitted through the device simultaneously without any interaction or creation of spurious signals. Due to the wave-based approach of computing in magnonic networks, the technique of frequency-division multiplexing can be the basis for parallel data processing in single magnonic devices, enabling the multiplication of the data throughput.

Main text:

Multiplexing is a widely used technique of data transmission in telecommunication or computer networks.[1] The common aim of all different realizations of this concept is to transfer multiple data signals through a shared transmission line.[2–6] In view of the different challenges todays CMOS technology is facing,[7] the concept of multiplexing is also very interesting for new approaches of data processing like, e.g., the field of magnonics.[8–14] In this case, spin waves (SW) are used to transport data[15-17] and the information can be encoded in their amplitude or phase.[18] This approach is especially interesting since wave-based logic can be realized by utilizing interference effects[19] of the spin waves as has been shown by, e.g., the development of the spin-wave majority gate[20,21] and other interferometer-based devices.[22,23]

Aiming at an efficient signal transport in magnonic circuits, spin-wave multiplexers enabling time-division multiplexing have already been experimentally demonstrated.[24,25] In this case, the information carrying spin-wave signals are getting temporarily separated, successively transferred through the shared magnonic waveguide and finally allocated to different output waveguides by the spin-wave (de)multiplexer.

In contrast, frequency-division multiplexing (FDM) enables a simultaneous transport of data.[1,26] The bandwidth of the transmission medium is divided into separated frequency channels and the data signals are simultaneously transferred at different frequencies. This approach is heavily used in a broad range of applications, from radio broadcasting and fiber optics to communication satellites.[3–6] However, since the data handling is realized by transistor-based logic elements, the information cannot be simultaneously processed in a single device. The data has to be demodulated and divided, since one transistor can only process one bit at a time.[26]

One significant advantage of wave-based logic (including spin waves) arises from the fact that single logic elements can perform multiple data operations simultaneously.[26] This is due to the superposition principle enabling independent interference-based operations in the different frequency channels. Hence, parallel data processing in single logic elements, in addition to simultaneous data transport, makes FDM especially interesting for wave-based computing since it multiplies the throughput of the logic networks. As important preparatory work, various approaches for the realization of a frequency-selective propagation of spin waves have already been demonstrated.[27-30]

We propose to utilize caustic-like spin-wave beams to realize FDM in magnonic circuits since they exhibit a pronounced frequency dependence of their propagation direction.[31-34] The origin of these beams is the anisotropic spin-wave propagation in an in-plane magnetized 2D magnetic medium which is induced by the dipole-dipole interaction. Considering a fixed spin-wave frequency, this anisotropy can lead to a (nearly) unchanged direction of the group velocity vector for a broad range of wavevectors. In this case, a focusing of spin-wave energy into designated propagation directions occurs which results in the creation of strong spin-wave beams.[32-39] A change of the spin-wave frequency leads to modified directions of the group velocity vector and, subsequently, to a change of the focused beam direction, as has been experimentally demonstrated.[32-35] We refer to all of these focused spin-wave beams, whose origin lies in the anisotropy of the spin-wave system, as "caustic-like". This is due to the fact that a spin-wave caustic in the narrower sense only occurs at the point of the constant-frequency curve in $k$-space where its curvature is zero[31]. This condition is not necessarily fulfilled for all the beams studied in this work. However, the following simulations as well as previous experiments[33,34] show no clear distinguishing features between the caustic-like and true caustic beams.

In this letter, we present the design of a passive device which performs frequency-division multiplexing in a microscaled magnonic logic network based on caustic-like spin-wave beams. As groundwork, the frequency dependence of the propagation direction of these beams is studied by micromagnetic simulations using the open-source simulation program MuMax3.[41] Subsequently, it is demonstrated how this effect can be used to combine and split spin-wave signals of different frequencies resulting in the developed FDM device. Finally, the frequency spectra in the different component parts of the FDM circuit are studied to verify the proper separation of the frequency channels. It is shown that in the low-signal regime no spurious signals due to, e.g., nonlinear frequency mixing occur and the spin-wave phase is preserved enabling its use for data coding.

In magnonic networks, the interconnections between logic elements are realized via narrow stripes of magnetic material, so called spin-wave waveguides. Hence, it is very suitable to perform the creation of caustic-like spin-wave beams within these networks by attaching an input waveguide to a widened film area.[34,36-38] Based on this approach, the geometry to study the frequency dependence of caustic-like spin-wave beams in microstructured ferromagnets is chosen as shown in **Figure 1a**. In all following cases, the material parameters of the Heusler compound $Co_2Mn_{0.6}Fe_{0.4}Si$ (CMFS[33,42-46]) have been used since this promising magnonic material combines high saturation magnetization[44] with rather low magnetic damping at microwave frequencies[44,45] and, thus, allows for a long-distance spin-wave propagation.[46] Due to the high saturation magnetization $M_S$, the efficient creation of caustic-like spin-wave beams is feasible in a broad range of frequencies,[33] which is not necessarily the case for materials with lower $M_S$ such as, e.g., YIG.[32] The 30 nm thick structures are in-plane magnetized along the y-direction by an external magnetic field of $\mu_0 H_{ext}$ = 75 mT. In the micromagnetic simulations, spin waves, which represent the input signal, are excited in the 1.5 µm wide waveguide at the position $x = -3$ µm by a locally applied microwave magnetic field. Figure 1b shows the result of the simulations for a spin-wave frequency of $f$ = 9.5GHz in the form of a two-dimensional map of the time-averaged, frequency-dependent spin-wave intensity (see experimental section for further information). Spin waves are excited in the input waveguide. The sharp transition from the waveguide to the broad magnetic area ($x = 0$ µm) acts as a point-like source for secondary spin waves exhibiting a broad wavevector spectrum, which leads to the creation of caustic-like spin-wave beams as discussed above. In contrast to previous works,[34,36-38] the input waveguide is positioned on the edge of the widened area. The boundary of the magnetic structure leads to a reflection of spin-wave energy rather than to its total absorption.[36,47,48] Thus, the geometric confinement prevents from a splitting of the spin-wave energy into two symmetric beams and only one caustic-like beam occurs. Its angle of propagation $\theta$ is defined with regard to the direction of the local magnetization,[36,38] which coincides with the direction of the external field inside the widened film area. To determine the propagation angle of the beam, a linear curve fitting has been performed along the position of the intensity maxima between $x = 1$ µm and $x = 11$ µm. It reveals a value of $\theta = (77.5 \pm 1.0)°$ with respect to the external magnetic field $H_{ext}$. To study the frequency dependence of the beam direction, this procedure has been repeated in steps of 0.5 GHz for frequencies between $f = 9$ GHz and $f = 18$ GHz. The lower limit is determined by the bottom of the spin-wave spectrum, which can be calculated to $f_b = (8.92 \pm 0.05)$ GHz.[49,50] The graph of Figure 1c summarizes the result of this study by showing the occurring beam angles $\theta$ as a function of the excitation frequency $f$. A clear saturation-like behavior is discernible: The changes of the beam angle are most pronounced close to the minimal spin-wave frequency $f_b$ and, within the studied frequency range, the angle variation is diminishing as $f$ increases. This behavior results from the excited wavevector spectrum in combination with the wavevector dependent direction of the group velocity vector at different frequencies which is determined by a complex interplay between the dipole-dipole and the exchange interaction: The dipole-dipole interaction induces the anisotropy of the spin-wave propagation in in-plane magnetized films and, hence, it is responsible for broad ranges of spin-wave wavevectors with (nearly) parallel directions of the corresponding group velocity vectors. Thus, the dipole-dipole interaction is mainly responsible for the creation of caustic-like spin-wave beams in this work. However, the exchange interaction influences the exact direction of the group velocity vectors and, hence, the studied frequency dependence of the direction of the caustic-like spin-wave beams. It should be mentioned that a creation of caustic-like spin-wave beams is also possible in the exchange dominated regime of the spin-wave spectrum.[52] Hence, a miniaturization of the presented multiplexing device utilizing spin waves with extremely short wavelengths should be possible. Aiming at an approach to realize frequency-division multiplexing in a magnonic logic circuit, the strong frequency dependence of the beam directions in the vicinity of the bottom of the spin-wave spectrum has to be highlighted. It becomes particularly obvious when comparing, e.g., the beam angles $\theta_{B1}$(9.5 GHz) = 77.5° and $\theta_{B2}$(12.5 GHz) = 66.2°. Hence, a frequency variation of $\Delta f$ = 3 GHz results in a pronounced change of the beam angle of $\Delta\theta_B$ = 11.3°. The distinct angle variation demonstrates the applicability of the caustic effect to separate and combine spin-wave signals of different frequencies.

Based on these insights, the design of a FDM prototype circuit has been developed, which is adjusted for the above mentioned frequencies $f_1$ = 9.5 GHz and $f_2$ = 12.5 GHz. The magnetic structure is indicated by the surrounding black, dashed lines in **Figure 2**. The circuit consists of three main component parts: The first part of the structure acts as a *combiner*, also referred to as *multiplexer*. Two 1.5 µm wide input waveguides assigned to two different input frequencies $f_1$ and $f_2$ are connected to the widened film area. The spike-shaped structure between the two inputs suppresses the spreading of spin-wave energy from the upper input waveguide into the negative *y*-direction during the beam formation. At the end of the 2D combiner area, a single output is added so that the energy of both inputs can be collected if their corresponding spin-wave beams occur under the proper angles. This 1.5 µm wide center waveguide constitutes the shared transmission line for the combined signals and, thus, is representing the second and main building block of the circuit, the actual *multiplex area*. After the pass-through of the spin waves, it also acts as the input for the third component part. The spin-wave *demultiplexer* separates the spin-wave signals of different frequencies again. It consists of another widened film area and two 1.5 µm wide output waveguides. The transition zones of the outputs of both the combiner and the demultiplexer are specially tailored and their positions are adjusted to ensure an efficient channeling of spin-wave energy into the subsequent waveguides according to the occurring beam directions.[38]

In the micromagnetic simulations, a simultaneous excitation of spin waves of different frequencies takes place in both input waveguides. This is realized by applying the microwave magnetic field $b_{\text{exc}}$, which is described in the experimental section, at the position $x = -3$ µm and centered in the respective input waveguide. In the upper input, a frequency of $f_1 = 9.5$ GHz is used while the excitation in the lower one is done at $f_2 = 12.5$ GHz. Figure 2a and b show the normalized spin-wave intensity distribution in the circuit at the respective frequencies. It should be stressed that the different input signals propagate through the structure at the same time. The separation of the frequency-dependent spin-wave intensities shown in Figure 2 was achieved by the employed temporal Fourier-transformation (see experimental section).

Exemplarily, the passage of the spin-wave signal of frequency $f_1 = 9.5$ GHz through the circuit will be discussed (Figure 2a). The spin waves enter the device in the upper input waveguide. In the combiner region, caustic-like spin-wave beams are created and they reach the multiplex area due to their propagation angle of $\theta_{B1}(9.5 \text{ GHz}) = 77.5°$. After passing this center waveguide, the signal enters the demultiplexer and is channeled back to the upper output. Following the same principle, spin waves of the frequency $f_2 = 12.5$ GHz entering the lower input are channeled into the same multiplex area and finally guided into the lower output again (Figure 2b). In agreement with Ref.,[38] the caustic-like spin-wave beams, which propagate in the combiner and demultiplexer, are transferred back to guided waveguide modes in the multiplex waveguide and in the outputs. Hence, this design of a FDM circuit makes it possible to combine two spin-wave signals of two different frequencies into a shared transmission line and to separate them again afterwards. It should be highlighted that this device is based on intrinsic effects of the creation of caustic-like spin-wave beams and no external control, such as, e.g., utilized in Ref.[24,25] to realize time-division multiplexers, is required.

After demonstrating the working principle of the FDM circuit, it will be verified that both spin-wave signals with different frequencies can pass the device simultaneously without disturbing each other by, e.g., nonlinear scattering effects. In the following paragraph, it will be shown that such effects, which might lead to spurious signals at other frequencies followed by an intensity decrease of the actual signals, are negligible.

Utilizing the data calculated by the temporal Fourier-transformation as described in the experimental section, **Figure 3**a-e shows the frequency spectra of the spin-wave amplitude at five different positions within the FDM circuit.[51] They are located in the center of the different waveguides as marked by the dots in Figure 2. The solid lines in Figure 3 represent the results in case a simultaneous excitation of both input signals takes place. In addition, simulations with single frequency excitations in the respective input waveguides have been performed as shown by the dashed lines. At each position, the data curves are normalized to the maximal amplitude of all three different excitation schemes.

In the case of a single frequency excitation at $f_1 = 9.5$ GHz in the upper input waveguide (red dashed lines), only one sharp peak at $f_1$ is visible in the upper in- and output as well as in the multiplex area (Figure 3a,c and d). The ratio of the signal at $f_1$ in the upper waveguides to the noise amplitudes in the lower waveguides is always higher than 17 dB. Hence, only very small signals are detected in the lower in- and outputs (Figure 3b and e). A spin-wave excitation in the lower input at $f_2 = 12.5$ GHz leads to a qualitatively equal observation with, however, dominant spin-wave signals at $f_2$ in the lower in- and output instead. This result highlights the efficient channeling and separation of the spin-wave signals into the respective waveguides once more.

In the case of a simultaneous excitation of spin-waves at $f_1$ and $f_2$ in the respective inputs, the resulting spectra are the linear combination of the previously observed spectra of the single frequency excitations and no intensity decrease of the maxima occurs. Inside the in- and outputs, only one pronounced peak is visible and two distinctly detached peaks occur in the multiplex area. Hence, two important facts are revealed. First, the spin-wave signals are transmitted in two clearly separated frequency channels as can be seen in Figure 3c. Second, the discrete peaks in the outputs in the case of a simultaneous transmission coincide in form and also in the maximum value with the curves which are observed in case of single frequency excitations. In all cases, the signal-to-noise ratio in the outputs between signals at $f_1$ and $f_2$ is still higher than 17 dB. No pronounced signals at other frequencies appear. This demonstrates that the two spin-waves at $f_1$ and $f_2$ do not interact inside the multiplex area and no frequency mixing occurs. The simultaneous propagation without interaction can be explained by the superposition principle in combination with the fact that the spin-wave amplitudes are in the linear regime. In the case of higher signal amplitudes, nonlinear phenomena might occur leading to scattering and mixing effects resulting in the creation of spin-waves at other frequencies.[45]

In addition, it has been verified that the phase of the spin-waves is not changed in case a simultaneous transmission takes place instead of single frequency excitations. This is shown in Figure 3f-e by comparing the phase evolution of the spin-wave amplitudes integrated along the waveguide width in the output at $t = 12.5$ ns.[38] Furthermore, the shown spatial phase evolution confirms a defined propagation of guided modes in the outputs.[38] A change of the transversal mode structure[53,54] during the propagation of the spin waves through the different waveguides of the presented circuit is not observed.

In conclusion, the presented circuit is a passive network element which performs the operation of multiplexing and demultiplexing of spin-wave signals without any additional power consumption. The studied frequency dependence of the direction of caustic-like spin-wave beams is used to combine and separate spin-wave signals of different frequencies enabling frequency-division multiplexing for simultaneous signal transport. It is evidenced that no frequency-mixing or other nonlinear effects occur in the shared transmission line and, thus, no spurious signals disturb the functionality. In addition, the signal phase is not affected in case of a simultaneous transmission. By inserting proper logic elements in the multiplex area, simultaneous data processing in a single device can be realized resulting in a large efficiency boost of the logic circuits. Hence, the presented device can be a pivotal element for multi-frequency magnonic logic networks.

**Experimental Section**

*Micromagnetic Simulations*: The micromagnetic simulations presented in this paper have been carried out by utilizing the open-source simulation program MuMax3.[41] In all cases, the magnetic structures are discretized into cells of 15 nm × 15 nm × 30 nm ensuring that in-plane wavevectors up to at least 100 rad/µm can be resolved. To suppress reflections of spin-wave energy at the outer boundaries of the simulated geometry, the Gilbert damping parameter is incrementally increased in 25 steps over a distance of 0.5 µm from $\alpha = 0.0021$ to $\alpha = 0.5$ at the outermost left edge of the input waveguides and the outermost right edge of the magnetic structures. As the first step of the micromagnetic simulations, the magnetic ground state is calculated. Afterwards, spin waves are excited by applying a local excitation field. It is oriented along the *z*-axis and has a Gaussian shape of the form $b_{\mathrm{exc}} = (0,0,b_0) \exp[-(x/0.2\ \mu\mathrm{m})^2 - (y/0.4\ \mu\mathrm{m})^2] \sin(2\pi f t)$ with an amplitude of $b_0 = 1$ mT. This profile ensures the possibility to excite spin waves in a broad range of wavevectors $k_x$ along the *x*-direction. In addition, the first transversal waveguide mode is dominantly excited due to the Gaussian profile in *y*-direction. Both features of the excited spin waves are desired for the utilization in a magnonic logic network. A maximum precession angle of the magnetization of around 0.45° occurs in the center of the excitation field. The simulation time is adjusted in accordance with the extend of the magnetic structures to ensure a propagation of the spin-wave energy through the whole geometry. In case of the results presented in Figure 1, the simulation time is set to 7.5 ns. Due to the bigger extent of the FDM circuit shown in Figure 2, the simulation time is increased to 15 ns. The resulting raw data of the micromagnetic simulations is the spatially resolved magnetization distribution. During the whole simulation time, the 2D data is saved every 25 ps right after applying the excitation field. In Figure 1b and 2, the results are presented in the form of time-averaged, frequency-dependent spin-wave intensity maps which have been obtained by a temporal Fourier-transformation and an integration of the resulting frequency-dependent spin-wave intensity over an interval of [$f - 0.25$ GHz, $f + 0.25$ GHz]. By employing the temporal Fourier-transformation, a separation and an independent discussion of the multi-frequency spin-wave signals, which propagate simultaneously through the presented multiplexing circuit, is possible. Furthermore, the results of the calculation provide access to the time-averaged, frequency-dependent spin-wave intensity which can be experimentally measured by Brillouin light scattering spectroscopy.[55] The normalized intensity shown in Figure 2 is calculated by normalizing the spin-wave intensity along the *y*-direction to its sum. By this, the spin-wave damping is compensated to illustrate the working principle of the device. In addition, only the data of the stationary state (t > 10 ns) is used to calculate the frequency dependent spin-wave intensity distribution by performing the previously described evaluation steps (The spin waves reach the end of the output waveguides after around 10 ns). This is important to ensure a proper calculation of the frequency spectra shown in Figure 3. In general, MuMax3 is capable of studying non-linear effects of spin waves with high amplitudes.[56] However, we restrict our analysis to the spin-wave propagation in the linear regime. Consequently, no non-linear effects are visible in the presented simulations as discussed in the main article.

*Design of the multiplexing circuit*: The multiplexing circuit has been designed to enable an efficient spatial separation of spin-wave signals of different frequencies. This separation can mainly be influenced by two dimensions of the multiplexing circuit: the width of the waveguides and the length of the widened film areas of the combiner and the demultiplexer.

The width of the waveguides, which are connected to the combiner and to the demultiplexer, determines the wavevector spectrum of the spin waves in the widened areas. A change of the wavevector spectrum can influence the directions of the caustic-like spin-wave beams due to a changed distribution of its containing group velocity vectors. This could engender a change of the spatial separation of spin-wave beams of different frequencies. However, this effect is not very pronounced for moderate changes of the waveguide width since the condition for the creation of caustic-like spin-wave beams is precisely that the direction of the spin-wave group velocity vector is (nearly) unchanged for a broad range of spin-wave wavevectors.

The second influence on the separation of different spin-wave signals is more pronounced and mainly due to geometric reasons: To reach a sufficiently large spatial separation of the signals, the length of the widened areas has to be adjusted to the width of the waveguides and the angle difference between the different caustic-like spin-wave beams. The (initial) width of the beams is given by the width of transition zones. Hence, a decrease of the waveguide width in combination with an unchanged length of the widened film areas would lead to an enhanced spatial separation of the different caustic-like spin-wave beams. This statement is valid as long as the direction of the beams is not significantly influenced by the decreased waveguide width.

*Spin-wave transmission:* The ratio of the spin-wave intensities integrated along the waveguide width at $x = -2$ µm and $x = 46$ µm are calculated to 26.0 dB for the signal at 9.5 GHz (upper in- and output) and 29.8 dB for the signal at 12.5 GHz (lower in- and output). These values can be mainly attributed to the intrinsic spin-wave damping of the magnetic material in combination with the respective propagation distances and group velocities of both signals, as it is the case for all spin-wave based devices. Apart from that, reflections of the spin-wave signals lead to additional transmission losses. It should be mentioned that the frequency-selective technique of (localized) parallel parametric amplification[50] could be used to compensate for losses and to equalize varying intensities of spin-wave signals in a multi-frequency magnonic logic circuit.

*Calculation of the bottom of the spin-wave spectrum*: The calculations are based on Ref.[49,50] and take into account the material parameters of CMFS, the film thickness of 30 nm and the effective magnetic field of $\mu_0 H_{\text{eff}} = (73 \pm 1)$ mT, which considers demagnetizing fields inside the film area.

**Acknowledgements**

The authors thank A.V. Chumak, Q. Wang and T. Ludwig for valuable discussions. Financial support by DFG within project B01 of the Transregional Collaborative Research Centre (SFB/TRR) 173 "Spin+X" and from the EU Horizon 2020 research and innovation programme within the CHIRON project (contract number 801055) is gratefully acknowledged.

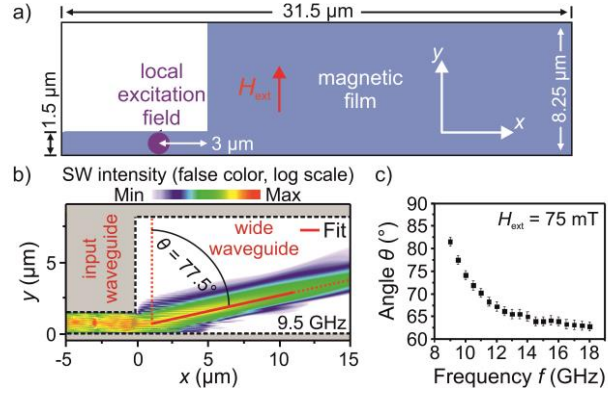

**Figure 1.** a) Sketch of the geometry used in the micromagnetic simulations. The magnetic structure consist of a narrow input waveguide connected to a wide film area. Spin waves are excited by a local excitation field inside the input waveguide. b) Resulting spin-wave intensity distribution for an excitation frequency of $f = 9.5$ GHz. A linear fit along the maxima of the occurring caustic-like spin-wave beam in the range from $x = 1$ µm to $x = 11$ µm is added. It reveals a propagation angle of $\theta = 77.5°$ with respect to the direction of the external field. c) Frequency dependency of the occurring beam directions.

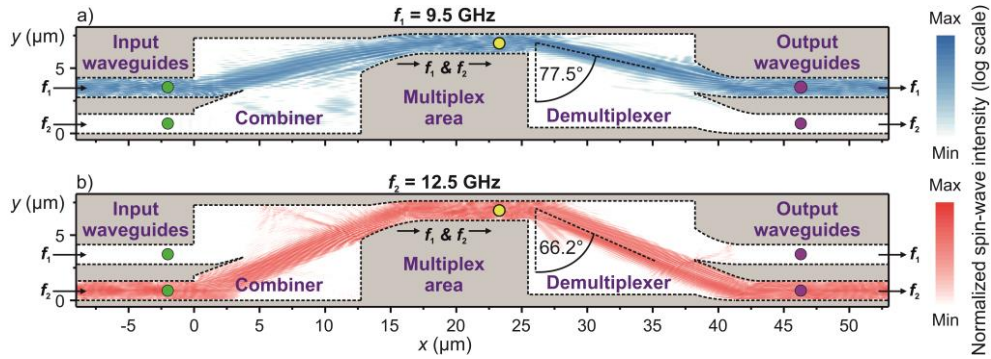

**Figure 2.** Frequency-division multiplexing circuit for magnonic logic networks based on caustic-like spin-wave beams. Spin-wave signals of different frequencies enter the circuit simultaneously through different input waveguides. Due to different beam angles, they are combined into the shared multiplex area and afterwards separated again into different output waveguides. The procedure is shown for spin-wave signals at a) $f_1 = 9.5$ GHz and b) $f_2 = 12.5$ GHz. The colored dots mark the extraction positions of the frequency spectra shown in Figure 3a-e.

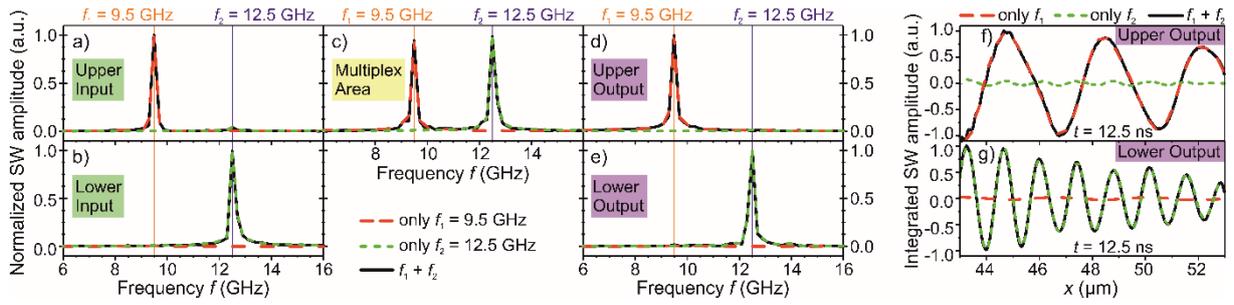

**Figure 3.** a) - e) Frequency spectra at different positions within the FDM circuit (marked by the dots in Figure 2). f) - g) Spin-wave intensity integrated along the width of the output waveguides as function of the *x*-position at $t = 12.5$ ns. All graphs show the results for three cases of spin-wave excitation: independent single frequency excitations at $f_1 = 9.5$ GHz and $f_2 = 12.5$ GHz in the respective waveguides (dashed lines) as well as simultaneous excitation of both frequencies $f_1$ and $f_2$ (solid lines).